 \documentclass[10pt,preprint]{aastex}  

\def\gapprox{\lower.4ex\hbox{$\;\buildrel >\over{\scriptstyle\sim}\;$}}
\def\lapprox{\lower.4ex\hbox{$\;\buildrel <\over{\scriptstyle\sim}\;$}}

\shortauthors{STURROCK AND ASCHWANDEN 2012}
\shorttitle{Crab Nebula Flares}

\begin{document}

\title{         Flares in the Crab Nebula Driven by Untwisting Magnetic Fields }

\author{        Peter Sturrock$^1$ and Markus J. Aschwanden$^2$		 }

\affil{         $^1$) Center of Space Science and Astrophysics,
		Stanford University, Stanford, CA 94305, USA;
		e-mail: sturrock@stanford.edu}
\affil{
		$^2$) Lockheed Martin Advanced Technology Center,
                Solar \& Astrophysics Laboratory,
                Org. ADBS, Bldg.252,
                3251 Hanover St.,
                Palo Alto, CA 94304, USA;
                e-mail: aschwanden@lmsal.com}

\begin{abstract}
The recent discovery of PeV electrons from the Crab nebula, produced 
on rapid time scales of one day or less with a sharply peaked gamma-ray
spectrum without hard X-rays, challenges traditional models of diffusive
shock acceleration followed by synchrotron radiation.
Here we outline an accleration model involving a DC electric
field parallel to the magnetic field in a twisted toroidal field
around the pulsar. Sudden developments of resistivity in localized 
regions of the twisted field are thought to drive the particle acceleration,
up to PeV energies, resulting in flares. 
This model can reproduce the observed time scales of $T \approx 1$ day,
the peak photon energies of $U_{\Phi,rr} \approx 1$ MeV, maximum electron
energies of $U_{e,rr} \approx 1$ PeV, and luminosities of $L \approx 10^{36}$ 
erg s$^{-1}$. 
\end{abstract}

\keywords{Pulsars --- particle acceleration --- magnetic fields}

\section{Introduction}

The {\sl Crab Nebula}, located at the center of the SN-1054 supernova
remnant, consists of a central pulsar with bi-directional jets and 
a relativistic particle wind (see Hester 2008 for a general review). 
While the optical, radio, and gamma-ray emissions originating from 
the pulsar and nebula vary only a few percent (the nebula is used 
as a standard candle), four most powerful flare episodes 
(with flux increases up to a factor of 6 times the background)
have recently been discovered with the AGILE satellite during October 
2007 and 19-21 September 2010 (Tavani et al.~2011, Abdo et al.~2011), 
and also by the {\sl Fermi Gamma-ray Space Telescope} 
during February 2009 and April 2011 (Buehler et al.~2010).

The following are some of the puzzling features of the gamma-ray flares 
that have recently been detected in the Crab Nebula:

\begin{enumerate}
\item{The photon energies are extraordinarily high, in excess of 1 GeV, 
	based on the gamma-ray emission observed by AGILE in the energy 
	range of 100 MeV $-$ 10 GeV (Tavani et al.~2011).}
\item{The timescale is quite short for an event that occurs in the nebula 
	- of the order of one day or less 
	(Tavani et al.~2011, Balbo et al.~2011, Striani et al.~2011). 
	The flares in February 2009 and September 2010 lasted about 16 days and 
	about 4 days (Abdo et al.~2011). The relative brevity of the 
	flares suggests gamma rays produced by gyrosynchrotron emission from
	$\approx 10^{15}$ eV (=PeV) electrons in a region $\lapprox 0.014$ 
	parsec (Abdo et al.~2011). The origin of variable multi-TeV 
	gamma-ray emission from the Crab nebula has been modeled in 
	Komissarov and Lyutikov (2011) and Bednarek and Idec (2011).}
\item{The spectrum rises steeply below the peak energy, which may be 
	from 7 MeV to 3 GeV.} 
\item{As a result, there is no detection of soft or hard X-rays associated 
	with the flares (Ferringo et al.~2010, Markwardt et al.~2010).}
\end{enumerate}

In this paper we develop a theoretical model of the PeV acceleration region,
in terms of twisted magnetic fields, which can explain the observed
extreme flare parameters within the right order of magnitude.

\section{Flare and Particle Acceleration Model}

For synchrotron radiation, the energy of the emitted photons (in eV) 
is given by,
\begin{equation}
	U_{\gamma} = 10^{-20.5} U_e^2 B \ ,
\end{equation} 	
where energies are measured in eV and magnetic fields strengths in Gauss.
	
Hence a photon of energy $U_\gamma = 1$ GeV is produced by an electron of 
energy $U_e = 10^{14.8} B^{-1/2}$.	
If the magnetic field strength is $B = 1$ Gauss or less, the electron energy 
will be $10^{15}$ eV or more. This raises two questions concerning the 
acceleration mechanism:

\begin{enumerate}
\item{How are electrons accelerated so rapidly to such a high energy?}
\item{Why is the electron spectrum so sharply peaked that it does not 
give rise to strong X-ray emission?}
\end{enumerate}

These puzzles suggest that we look for something different from 
the usual ideas of shock acceleration followed by ordinary synchrotron 
radiation.  We suggest that the acceleration is due to a process that 
develops a DC electric field parallel to the magnetic field, and that 
the radiation is due to the motion of electrons along curved magnetic 
field lines. It appears that these two processes can occur as a 
consequence of the sudden development of resistivity in a localized 
region of a twisted magnetic flux tube. 

We consider a pre-flare state that comprises a twisted flux tube 
held in place by a surrounding flux system. To bypass consideration of the 
termination of the tube, we suppose that it has the form of a toroid 
of major radius $R$ and minor radius $R_1$. The mean magnetic field strength 
is $B$, and the mean electron density is $n$ (cm$^{-3}$). 

If the mean value of the azimuthal or transverse magnetic field is $B_1$, 
then the total free magnetic energy is,
\begin{equation}
	W = 2 \pi R \pi R_1^2 {1 \over 8 \pi} B_1^2 
	  = 10^{-0.1} k_R^2 k_B^2 R^3 B^2 \ .	
\end{equation}
where we write,
\begin{equation}
	R_1 = k_R R \ , \quad
	B_1 = k_B B \ .
\end{equation}
Now suppose that an instability (probably a two-stream instability) 
occurs somewhere along the length of the toroidal flux tube, and that 
the effect of this instability is equivalent to suddenly inserting 
a highly resistive sheet across the flux tube. 

If $n$ is the electron density, the relativistically correct expression 
for the Alfven speed is (Sturrock 1994)  
\begin{equation}
	v_A = \left( {1 \over c^2} + {4 \pi n m_p \over B^2} \right)^{-1/2} \ ,
\end{equation}
where $m_p$ is the ion (proton or positron) mass.
In subsequent calculations we assume that $v_A \approx c$  to adequate 
approximation. 

The tube will begin to unwind, and the time it takes to unwind is	
\begin{equation}					
	T = {2 \pi R \over v_A} = {2 \pi R \over c} = 10^{-9.7} R \ .
\end{equation}
The total energy release rate (the total luminosity) is given by,
\begin{equation}
	L = {W \over T} = 10^{9.6} k_R^2 k_B^2 R^2 B^2 \ .
\end{equation}
This process leads to an electric field parallel to the magnetic field. 
We estimate the magnitude of this field from 				
\begin{equation}
	\nabla \times {\bf E} = - {1 \over c} {\partial {\bf B} \over 
	\partial t} \ ,
\end{equation}
where we use modified Gaussian units (electric units in esu, and magnetic 
units in emu). This leads to the estimate
\begin{equation}
	E = {B_1 R_1 \over c T} \approx 10^{-0.8} k_R k_B B \ .	   
\end{equation}
The maximum energy that can be imparted to an electron (one that travels 
the complete circuit) is 
\begin{equation}
	U_{e,max} = e E 2 \pi R = 10^{-9.3} k_R k_B R B \ .
\end{equation}
However the actual energy of an electron is likely to be 
radiation reaction-limited. Electrons of energy $U_e$ (eV) traveling along 
a field line with radius of curvature $R$ will emit, by curvature radiation, 
photons of energy (in eV) 
\begin{equation}
	U_{\Phi} = 10^{-22.2} U_e^3 R^{-1} \ ,
\end{equation}
and the rate of radiation of energy per electron is given by 
\begin{equation}
	S_e = 10^{-31.1} U_e^4 R^{-2} \ .
\end{equation}
(Note however that the energy released due to the change in the magnetic 
field could go into more than one form of energy - the energy converted to 
radiation, and the kinetic energy of the relaxed plasma.  Any excess released 
energy could set up an oscillatory macroscopic motion of the plasma plus 
magnetic field.)      

On equating the rate of energy input to the rate of energy output,	
\begin{equation}
	e c E = 10^{0.4} k_R k_B B = 10^{-31.1} U_e^4 R^{-2} \ ,
\end{equation}
we arrive at the radiation reaction-limited electron energy (probably 
an estimate of the maximum electron energy): 	
\begin{equation}
	U_{e,rr} = 10^{7.9} k_R^{1/4} k_B^{1/4} R^{1/2} B^{1/4} \ .
\end{equation}		
This leads to the following estimate of the photon energy:
\begin{equation}
	U_{\Phi,rr} = 10^{1.5} k_R^{3/4} k_B^{3/4} R^{1/2} B^{3/4} \ .
\end{equation}
However (assuming that the electron energy is radiation-reaction-limited), 
the total luminosity is also expressible as
\begin{equation}
	L = ( 2 \pi R \pi R_1^2 f n ) \times e c E \ ,		
\end{equation}
where $f$ is the fraction of the  available electrons that are actually 
accelerated.

This is found to be 
\begin{equation}
	L = 10^{1.7} k_R^3 k_B f n R^3 B \ .				
\end{equation}
On comparing Eq.~(16) with Eq.~(6), we obtain the following estimate for $f$,
\begin{equation}
	f = 10^{7.9} k_R^{-1} k_B n^{-1} R^{-1} B \ .
\end{equation}
To see if these formulas lead to 
reasonable estimates, we consider one of the flares that occurred 
in April 2011, for which we have the following observational constraints:

\begin{enumerate}
\item{The time-scale was reported as ``hours''. We adopt 10 hours, 
	so $T=10^{4.6}$.}
\item{The peak energy of one of the flares was reported to be 
	about 100 MeV so $U_{\phi,rr}=10^8$ eV.}
\item{The luminosity was estimated to be $L = 10^{36}$ erg s$^{-1}$.}
\end{enumerate}
Then Eq.~(5) leads to the radius $R = 10^{14.3}$ cm.
If we set $k_R = k_B = k$, then with this value of $R$, Eq.~(6) leads to 
$k^2 B = 10^{-1.1}$, and Eq.~(14) leads to $k^2 B = 10^{-0.9}$.
It is interesting that we can satisfy both equations to good 
approximation by adopting $k^2 B = 10^{-1}$. Then $k = 10^{-1}$
leads to $B = 10$ G, for instance. 

For these values, Eq.~(17) leads to $f n = 10^{-5.4}$. If  $n=10^{-5}$, 
then practically all of the electrons are accelerated; 
if $n=1$, then only a small fraction are accelerated, etc. 

There is really a lot of flexibility in these estimates, 
of course, but the above calculations show that the model can fit 
observational data with parameters that are not unreasonable. 

Concerning the luminosity, 
we should bear in mind that the radiation from any part of the toroid will 
be sharply beamed. If the flux tube were strictly toroidal in form, 
the radiation would have a disk-like polar diagram. Since the radiation 
from any location is beamed, the beam probably varies in time, so that (a) 
the inferred luminosity may be an overestimate of the true  luminosity, and 
(b) the observed duration of a flare may be an underestimate of its true 
duration. 

Finally, we comment briefly on a possible process for creating 
the sudden resistivity. The current requires that electrons and ions 
(or positrons) flow in opposite directions around the toroid. Suppose that, 
for some reason, the density locally becomes low enough that the relative 
flow velocities lead to a two-stream instability. This will give rise to 
local energy dissipation, hence to heating of that region of the plasma. 
But heating will lead to expansion, which will further reduce the density. 
In order to maintain the current, the velocity differential must increase, 
which will enhance the two-stream instability. Hence this combination of 
microscopic and macroscopic processes may lead to a sudden very strong 
two-stream instability that would make the region locally ``electrically 
turbulent'' and resistive. 

Why curvature radiation rather than conventional 
synchrotron radiation? The DC electric field drives electrons parallel 
to the magnetic field, so there is initially no transverse momentum to 
lead to synchrotron radiation. Furthermore, the gyro-radius of the 
electrons is much smaller than the radius of the toroid, so electrons 
will continue to follow field lines. For instance, for the parameters 
considered above, $T = 10^{4.6}$ s , $R = 10^{14.3}$ cm , $B = 10$ G, 
$k = 0.1$ we find that $U_{e,rr} = 10^{14.8}$ eV. The gyro radius 
would be given by  
\begin{equation}
	r_g = 10^{-2.5} E_e B^{-1} \ ,
\end{equation}
where $E_e$ is in eV and $B$ is in Gauss, if the electron 
velocity were transverse to the magnetic field. Hence we 
obtain the estimate $r_g = 10^{11.3}$ cm. Since this is much smaller 
than $R$, we may infer that
electrons will be moving essentially along the magnetic field lines. 
Hence radiation will be curvature radiation rather than conventional 
synchrotron radiation.

In Table 1 we show the model parameters for a set of 9 parameter sets,
where we choose $k=0.1$ in all models, magnetic field strengths of
$B=10^{-1}$ G (model A,D,G), $B=1$ G (model B,E,H), 
$B=10$ G (model C,F,I), and the length and radius of curvature
$L=10^{14}$ (model A,B,C), $L=10^{15}$ (model D,E,F), and 
$L=10^{16}$ (model G,H,I). For this parameter space we find 
energy release times or flare durations in the range $T=10^{3.5}$ s 
($\approx 1$ hour) to $T=10^{5.5}$ s ($\approx 3.6$ days),
electron energies limited by radiation reaction of 
$U_{e,rr} \approx 10^{14}-10^{16}$ erg (0.1-10 PeV), 
and corresponding photon energies 
$U_{\Phi,rr} \approx 10^7-10^9$ eV. Thus we can satisfy the observed
measurements within a wide range of the parameter space.

\section{Magnetic field in Crab Flares}

We depict the radial dependence of the magnetic field strength $B(r)$
as a function of distance $r$ from the center of the pulsar in Fig.~1.
The Crab pulsar has a spin period of $P=30$ ms, which yields 
a light cyclinder radius of $r_{LC}=P c/2 \pi = 10^{8.2}$ cm.
The surface magnetic field strength is $B_0=10^{12.6}$ G,
and the radius of the pulsar is 12.5 km or $r_0=10^{6.1}$ cm. 
The magnetic field inside the light cylinder is a dipole field
and falls off with an $r^{-3}$ dependence, which yields a field strength of
\begin{equation}
	B_{LC}= B(r=r_{LC}) = B_0 \left( {r_{LC} \over r_0} \right)^{-3}
	= B_0 \left( {10^{8.2} \over 10^{6.1}} \right)^{-3}
	=10^{6.3} \ {\rm G} \ ,
\end{equation}
at the distance $r_{LC}$ of the light cyclinder. Beyond the light cylinder,
the field wraps up like an onion skin (in the ``striped wind'' region), 
falling off inversely with distance. Hence at distance $r$, it has 
a value of
\begin{equation}
	B(r) = B_{LC} \left( {r \over r_{LC}} \right)^{-1}
	     = 10^{14.5} r^{-1} \ .
\end{equation}
The magnetic field strength in the nebula is estimated to be 200 mG,
i.e., $B_N = B(r=r_N) = 10^{-3.7}$ G. The pulsar field will have dropped
to this value at a distance of $r_N=10^{18.2}$ cm, which is one third
of the observed radius ($10^{18.7}$ cm) of the nebula. Hence it seems reasonable
to adopt the above relationship between $B$ and $r$ (Fig.~1).
The zone of particle acceleration by untwisting magnetic fields lies,
according to our model estimates (Table 1), at a distance of
$r \approx 10^{13}-10^{15}$ cm from the center of the pulsar
(shown with a hatched region in Fig.~1), so it is located far away from
the pulsar and the light cyclinder ($R \gapprox 10^5 r_{LC}$), 
but close to the pulsar when compared to the Crab nebula 
diameter ($R \lapprox 10^{-4} r_N$), such as seen in a Hubble or Chandra
image. 

\section{Predicted Scaling Law}

Based on our model we can predict a scaling law between the observables
of the photon energy $U_{\Phi}$, total luminosity $L$, and flare duration
$T$. The radiation reaction-limited photon energy $U_{\Phi,rr}$ varies
according to Eq.~(14) with the radial distance $R$ and the magnetic
field $B$ as,
\begin{equation}
	U_{\Phi,rr} \propto R^{1/2} B^{3/4} \ .
\end{equation}
Combining the relationships of Eq.~(16) and (17) we find that the total 
luminosity varies with the observables $R$ and $B$ as,
\begin{equation}
	L \propto R^2 B^2 \ .
\end{equation}
The duration of a flare (or unwinding episode of the twisted magnetic
field) varies linearly with $R$ (Eq.~5),
\begin{equation}
	T \propto R \ .
\end{equation}
From these three relationships (Eqs.~21-23) we can eliminate $R$ and $B$
and arrive at the following scaling law for the observables $U_{\Phi}$,
$L$, and $T$,
\begin{equation}
	U_{\Phi} \propto L^{3/8} T^{-1/4} \ .
\end{equation}
Thus, if the total luminosity $L$ and the time scale $T$ of a flare are 
observed, the photon energy $U_{\Phi}$ can be predicted, or vice versa.
Such a scaling law could be tested in the future either for a statistical
sample of flares from the same pulsar, or among different pulsars. 

\section{Discussion and Conclusions}

Previous models of particle acceleration in pulsar and supernova remnant
environments include: 
(i) magnetohydrodynamical (MHD) models (Rees and Gunn 1974), 
Kennel and Coroniti 1984), (ii) diffusive shock acceleration 
(Drury 1983, Blandford and Eichler 1987), 
(iii) shock-drift acceleration (Kirk et al.~2000, Reville and Kirk 2010), 
(iv) magnetohydrodynamical instabilities (Komissarov and Lyubarsky 2004,
Del Zanna et al.~2004, Camus et al.~2009, Komissarov and Lyutikov 2011),
(v) acceleration due to absorption of ion cyclotron waves (Arons 2008),  
(vi) runaway acceleration (de Jager et al.~1996), 
(vii) magnetic reconnection in an AC striped wind,
and (viii) a linear electric accelerator operating in the electric fields 
of reconnection sites (Uzdensky et al.~2011). 
Difficulties with the diffusive shock acceleration model arise 
because the synchrotron cooling length is comparable with the Larmor radius
(since the gyroperiod is a lower limit for the acceleration time scale), 
for the strong electric fields that are required to compensate the 
radiation reaction (Abdo et al.~2011, Striani et al.~2011). 

Here we propose an alternative mechanism of DC electric field acceleration 
parallel the the magnetic field of untwisting flux tubes. The particle
dynamics in the acceleration region (at a distance of $R \approx 10^{13}
-10^{15}$ cm (Table 1 and Fig.~1) is completely dominated by the magnetic 
field, since the plasma-$\beta$ parameter is much less than unity (i.e., 
$\beta = n k_B T/ (B^2/8\pi) \approx 2.5 \times 10^{-8}$ for $n=1$ 
cm$^{-3}$, $T \approx 10^7$ K, and $B \approx 1$ G). A twisted toroidal 
magnetic field can naturally be generated for a number of reasons:
for instance, by reconnection in the magnetic ``onion skin'' of the
``striped-wind'' region.

The spin-down of a fast-rotating neutron star provides free energy 
for a toroidal field.  Occasional sporadic glitches in the rotational 
frequency (with typical amplitudes of $\Delta \Omega / \Omega \approx 
10^{-9}-10^{-6}$), thought to occur due to neutron starquakes in the 
stressed crust of the neutron star, introduce sudden changes in the 
co-rotating magnetic field, which may cause sudden topological changes.
A sudden change in the current along the toroidal magnetic 
field that accelerates both electrons and positrons (in opposite directions) 
may lead to a particle velocity distribution with the right (positive) sign 
of the curvature of the distribution function to trigger a two-stream 
instability.  The mechanism of a two-stream instability to generate 
toroidal magnetic fields in the magnetosphere of the Crab pulsar 
has been studied in Nanobashvili (2011). The two-stream instability may 
also be due to the resistivity-driven development of microscopic or 
macroscopic turbulence outlined in Section 2.  

In fact, giant pulses from the Crab pulsar up to 200 times the average 
pulse size with a powerlaw distribution have been observed in gamma rays 
and radio wavelengths during previous flaring episodes (Argyle and 
Gower 1972; Lundgren et al.~1995), indicating sporadic giant energy releases 
in a system governed by self-organized criticality (Young and Kenny 1996; 
Warzawski and Melatos 2008). Similar powerlaw distributions have been 
found for magnetospheric substorms, solar flares, stellar flares, 
soft gamma-ray repeaters, black hole objects, blazars, and cosmic rays
(Aschwanden 2011), which all are believed to be caused in one way or the
other by nonlinear energy dissipation events by magnetic reconnection 
in twisted, stressed, and nonpotential magnetic fields. 

A statistical model for the $\gamma$-ray variability of the Crab nebula
has been simulated for electrons that have maximum energies proportional
to the size of knots (Yuan et al.~2011), which can also be applied
to the size of untwisting magnetic field regions in our model, yielding a
similar accelerated electron spectrum. Future observations of flaring
episodes from the Crab nebula should yield statistics of the flare time
durations $T$, of the spatial size $S$ of flaring knots, the
total energy release $W$, and the luminosity $L=W/T$. Modeling of the 
twisted magnetic field $B({\bf r})$ together with the free energy and current
density contained in the (non-potential) azimuthal component $B_{\varphi}$,
could then provide testable scaling laws between the energies released
in flare events and the observed physical parameters. Furthermore,
our predicted scaling law between the maximum photon energy $U_{\Phi}$,
luminosity $L$, and flare duration $T$ (Eq.~24) may prove to be   
appropriate for multiple flares from pulsars.

\clearpage


\begin{deluxetable}{llrrrrrrrrr}
 \tabletypesize{\normalsize}
\tabletypesize{\footnotesize}
\tablecaption{Model parameters (given in log-${10}$ and cgs-units) 
of flares from Crab Nebula with radiation reaction and relativistic 
Alfv\'en speeds.}
\tablewidth{0pt}
\tablehead{
\colhead{Parameter}&
\colhead{}&
\colhead{A}&
\colhead{B}&
\colhead{C}&
\colhead{D}&
\colhead{E}&
\colhead{F}&
\colhead{G}&
\colhead{H}&
\colhead{I}}
\startdata
Azimuthal field ratio $B_\varphi/B$	& $k_B$	&
 -1.0 &  -1.0 &  -1.0 &  -1.0 &  -1.0 &  -1.0 &  -1.0 &  -1.0 &  -1.0 \\
Magnetic field strength 		& $B$ &
 -1.0 &   0.0 &   1.0 &  -1.0 &   0.0 &   1.0 &  -1.0 &   0.0 &   1.0 \\
Length and radius of curvature		& $L$ &
 14.0 &  14.0 &  14.0 &  15.0 &  15.0 &  15.0 &  16.0 &  16.0 &  16.0 \\
Radius of flux tube			& $R$ &
 13.0 &  13.0 &  13.0 &  14.0 &  14.0 &  14.0 &  15.0 &  15.0 &  15.0 \\
Electron density			& $n$ &
  0.0 &   0.0 &   0.0 &  -1.0 &  -1.0 &  -1.0 &  -2.0 &  -2.0 &  -2.0 \\
Free magnetic energy			& $W$ &
 35.1 &  37.1 &  39.1 &  38.1 &  40.1 &  42.1 &  41.1 &  43.1 &  45.1 \\
Energy release time			& $T$ &
  3.5 &   3.5 &   3.5 &   4.5 &   4.5 &   4.5 &   5.5 &   5.5 &   5.5 \\
Energy release rate			& $U$ &
 31.6 &  33.6 &  35.6 &  33.6 &  35.6 &  37.6 &  35.6 &  37.6 &  39.6 \\
Electric field strength in esu		& $E$ &
 -3.0 &  -2.0 &  -1.0 &  -3.0 &  -2.0 &  -1.0 &  -3.0 &  -2.0 &  -1.0 \\
Radiation reaction limit to electron energy & $U_{e,rr}$ &
 14.4 &  14.6 &  14.9 &  14.9 &  15.1 &  15.4 &  15.4 &  15.6 &  15.9 \\
Radiation reaction photon energy        & $U_{\Phi,rr}$ &
  6.9 &   7.6 &   8.4 &   7.4 &   8.1 &   8.9 &   7.9 &   8.6 &   9.4 \\
\enddata
\end{deluxetable}

\clearpage

\begin{figure}
\plotone{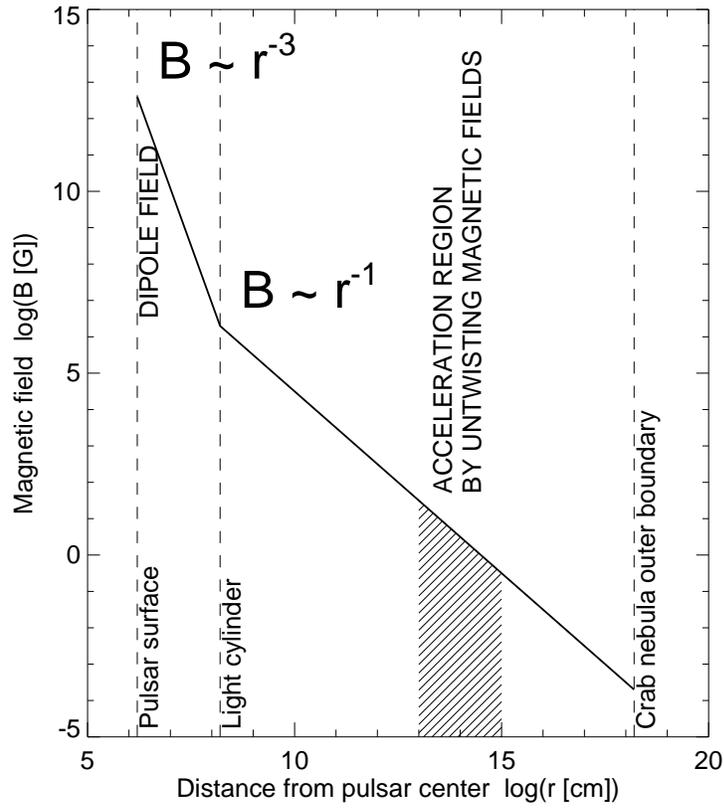}
\caption{Diagram indicating the dependence of the magnetic field $B(r)$ as a 
function of the distance $r$ from the center of the Crab pulsar, showing
the relative locations of the pulsar surface, the light cylinder, the
acceleration region by untwisting magnetic fields (according to our model),
and the Crab nebula.}
\end{figure}

\end{document}